# Geometrical Effect Explains Graphene Membrane Stiffening at Finite Vacancy Concentrations

Zhigong Song[1] and Zhiping Xu[1,2*]

[1]Applied Mechanics Laboratory, Department of Engineering Mechanics and Center for Nano and Micro Mechanics, Tsinghua University, Beijing 100084, China

[2]State Key Laboratory of Mechanics and Control of Mechanical Structures, Nanjing University of Aeronautics and Astronautics, Nanjing 210016, China

[*]Corresponding author, Email: xuzp@tsinghua.edu.cn

## Abstract

The presence of defects such as vacancies in solids has prominent effects on their mechanical properties. It not only modifies the stiffness and strength of materials, but also changes their morphologies. The latter effect is extremely significant for low-dimensional materials such as graphene. We show in this work that graphene swells while point defects such as vacancies are created at finite concentrations. The distorted geometry resulted from this areal expansion, in combination with the in-plane softening effect, predicts an unusual defect concentration dependence of stiffness measured for supported graphene membrane in nanoindentation tests, which explains the defect-induced stiffening phenomenon reported recently. The mechanism is elucidated through an analytical membrane model as well as numerical simulations at atomistic and continuum levels. In addition to elucidate the counter-intuitive observations in experiments and computer simulations, our findings also highlight the role of defect-modulated morphology engineering that can be quite effective in designing nanoscale material and structural applications.

## Keywords

Graphene, defects, geometrical effects, areal expansion, nanoindentation, defect-induced stiffening

The presence of defects in solids modifies their mechanical response, which is an important subject in understanding the structure-properties relationship of materials for long time.[1] For example, point defects not only reduce the material stiffness and strength due to the local mass deficiency and stress concentration they create, but could also strengthen materials through mechanisms of solid solution and precipitate hardening.[1] The role of defects in defining material properties is even richer when the dimension of materials is reduced. A number of studies have demonstrated prominent geometrical effects of topological defects in two-dimensional (2D) materials in additional to the stress field they create.[2-4] Although these materials are often loaded in plane in practical applications such as nanoelectromechanical devices and nanocomposites, their mechanical characterization are usually done through nanoindentation tests where concentrated load is applied towards a specific region of the material.[5-10] In this situation of local probes, the out-of-plane geometrical distortion could drastically change their mechanical response.[4]

Recently there were two individual studies reporting that finite-concentration of vacancies in the graphene membrane lead to a remarkable stiffening effect under nanoindentation probe, which is contrastive to the conventional understanding that point defects in materials usually soften their mechanical performance. However, the underlying mechanism has not been well clarified. In the work conducted by Kvashnin and co-workers,[11, 12] molecular dynamics (MD) simulation results are reported, showing that that the Young's modulus of the graphene membrane with the defects concentration up to $c = 1\%$ increases to 2.57 TPa. The further increasing of defects concentration ($c = 7\%$) leads to reduction in the stiffened value of Young's modulus to 1.08 TPa. The underlying mechanism is explained by a competition of two phenomena, which are the 'hardening' of the graphene membrane due to the shortening of bonds and reduction in the density of graphene lattice due to the presence of vacancies. However, the shortening in bond lengths near the defects in graphene may not necessarily stiffen the local atomic structures compared to the perfectly hexagonal lattice of graphene, and the effect is contrastive with our previous calculations when graphene is loaded in its basal plane.[11, 13] In their simulation setup, an indenter consisting of a few atoms are used to press the center of the supported graphene sheet. The 2D membrane stiffness $E_{2D}$ is calculated by fitting the indentation force-depth relation through the Schwering type of solution

$$f(\delta) = \pi\sigma_0\delta + E_{2D}\delta^3/a^2 \qquad (1)$$

where $f$ is the indentation force, $\delta$ is the indentation depth, $a$ is the radius of suspended graphene membrane, and $\sigma_0$ is the prestress. It should be remarked here that this solution is based on the assumption that the graphene sheet is considered as a planar membrane experiencing small-deflection deformation. However, a prestrain could be introduced due to the fact that graphene swells as vacancies are created,[18] and there will be an additional geometrical effect from this expansion – the graphene would buckle out of the plane. As a result, the membrane could suffer significant out-of-plane distortion under indentation, and the planar assumption could break down.

In another recent report,[5] results from nanoindentation experiments are presented, showing the same phenomenon of defect-induced stiffening, although the indentation strength is reduced that is in consistence with conventional understandings. The determination of elastic modulus is also achieved by fitting the experimental data from nanoindentation tests through **Eq. 1**. Similar results were also reported in Ref. 6. The mechanism for the counter-intuitive defect-induced stiffening effect is discussed based on an argument of thermal-fluctuation-induced wrinkles in the graphene membrane, through the dependence of the elastic coefficients on the momentum of flexural modes of two-dimensional membranes.[5] The key idea is that with wrinkles the membrane is softened,[9] while the introduction of vacancies will pin the long-wave fluctuations and thus effectively 'stiffen' the membrane compared with the membrane in absence of vacancies. However, this phenomenon-based reasoning is inconsistent with the abovementioned study using MD simulations[11] if we assume these two similar results arise from the same origin. In the MD simulations,[11] the formation of thermal wrinkles in graphene is negligible due to the limited lateral span of the supported membrane that is 3.8 – 14 nm, much smaller than the value of 0.5 – 3 μm in the nanoindentation experiments. Moreover, the argument of fluctuation renormalized tensile stiffness is valid only at thermal equilibrium, while by fitting the measured nanoindentation force-depth relation using **Eq. 1**, the second term that is responsible for extracting $E_{2D}$ only dominates at the large-displacement regime, where the membrane is stretched significantly, and the winkled structures are reduced or even disappear. Actually previous experimental work using nanoindentation tests measured reasonable stiffness for monolayer graphene where defects such as grain boundaries are present.[7-10] That is to say, the thermal fluctuation induced stiffness renormalization picture may not apply in these measurements. This inconsistence and lack of understanding for the counter-intuitive phenomenon drives us to explore the physical mechanism behinds the observed defect-induced stiffening effect. In the following text, we will first discuss the vacancy-induced in-plane softening and swelling behavior, and then elucidate the mechanism of stiffening effect of defective graphene membrane in response to nanoindentation loads, where we find a prominent geometrical effect that makes the key contribution.

In this work, we find that after mono-vacancies are introduced to the graphene membrane by exotic treatments such as irradiation, two major effects are introduced, which are critical for the subsequent mechanical characterization under nanoindentation. Firstly, the in-plane stiffness and strength are reduced upon defect creation.[13] Secondly, the defective lattice will expand that drives the membrane to buckle out of the plane due to the distortion of $sp^2$ bonding network, which stiffens the nanoindentation response due to a prominent geometrical effect as the membrane deforms significantly from a planar shape.[4, 14] These two factors compete with each other in modifying the stiffness and will dominant at high and low concentration of vacancies, respectively. As a result, there is a peak value of the two-dimensional elastic stiffness or the Young's modulus, which is extracted by fitting the nanoindentation results using **Eq. 1**.

To demonstrate this point, we firstly carry out full-atom MD simulations to probe the in-plane stiffness reduction and expansion upon the creation of mono-vacancies. In this work, MD simulations are performed using the large-scale atomic/molecular massively parallel simulator (LAMMPS) package.[15] The adaptive intermolecular reactive empirical bond-order (AIREBO) potential function is used to describe the interatomic interaction in graphene.[16] The monovacancy defects are created by randomly picking single carbon atoms with three bonded neighbors for removal in the graphene sheet. Although other types of defects such as divacancies and nanoholes could present in graphene under irradiation, we focus here on monovacancies that have been widely identified in experiments.[17] It should also be noted that at low temperature ~300 K, the mobility of vacancies is prohibited and thus the spatial distribution of defects is fixed during the mechanical tests.[17] 2D periodic boundary conditions (PBCs) are applied with a simulation box size of 33.5 x 36.3 nm. Both atomic positions and the box size are optimized to obtain stress-free relaxed structures of graphene sheets with embedded defects, using the conjugated gradient algorithm. To quantify the in-plane elasticity of defective graphene, a planar graphene sheet is then stretched uniaxially in its basal plane, and the Young's modulus is calculated by fitting the small-deformation tensile stress-strain curve using a linear function $\sigma = Y\varepsilon$. The results are summarized in **Fig. 1a**, which demonstrate an almost linear dependence between the reduction in in-plane Young's modulus and defect concentration $c$ that could be fitted as $Y = (-7.227c + 1.006)$ TPa. By measuring changes in the area of defective graphene, we also find a linear dependence between in-plane line expansion $\varepsilon_d = \Delta l/l$ due to the presence of monovacancies and $c$ at low defect concentration (**Fig. 1b**). Through a linear fitting, we have $\varepsilon_d = 0.167c$ for $c < 2\%$. We then define a coefficient of mono-vacancy-induced linear expansion as $r = \varepsilon_d/c$ at low concentration. Although these results are obtained by using empirical interatomic potentials, we verify its reliability by directly comparing the linear expansion coefficient $r = \varepsilon_d/c$ and the distortion of local atomic structure by the presence of defects with the results obtained from density functional theory (DFT) based first-principles calculations (see **Figs. S2**, **S3** and **Supplementary Information** for details). Such an evidence of areal expansion upon monovacancy creation is also consistent with recent experimental measurements where strain of expansion in graphene subjected to oxygen etching and ion bombardment could reach 2.2% for graphene on Ir and 0.3% for graphene on $SiO_2$,[18] as well as first-principles calculations.[19, 20] Moreover, based on the local lattice distortion calculated from either MD or DFT, one could calculate the linear expansion coefficient of areal expansion. Our estimation following this approach aligns with the direct MD simulations shown in **Fig. 1b** (more details can be found in the **Supplementary Information**).

In the nanoindentation tests, graphene sheets are supported on porous substrate (**Fig. 2**). The interfacial adhesion between graphene and the substrate constrains the in-plane deformation in the non-suspended region. As a result, the creation of vacancies in graphene does soften the in-plane elastic modulus. On the other hand, a notable areal expansion in the membrane is also induced during this process, which could lead to significant out-of-plane distortion or buckling as the membrane is constrained at the boundary. To characterize the structural distortion and mechanical response under the defect creation and indentation probe, we perform simulated nanoindentation tests on the suspended graphene membrane on a cylindrical pore with a diameter of $2a = 30$ nm. The setups are similar as that in recent experimental measurements[5-10] and our previous study of graphene with topological defects,[4, 14, 21] which is illustrated in **Fig. 2**.

We perform nanoindentation tests at the center of membrane with a downward speed of $v = 1$ m/s. A spherical indenter with radius of $R = 1$ nm is used. It should be noticed that the supported graphene sheets after structural relaxation show distinctly out-of-plane distortion because of both the defect-induced corrugation and in-plane expansion, which can then turns into vertical displacement with an amplitude of a few nanometers by unfolding the wrinkles after it makes contact with the indenter tip (**Fig. S1**). Our calculations show that a suspended graphene membrane with radius $a = 1$ μm experiences a out-of-plane displacement of 7.19 nm for $c = 2\%$, which aligns with the reported surface profile in Ref. 5, 6. During this unfolding process, the response in indentation force is small due to the low bending stiffness of graphene. Only if the wrinkles are flattened and the membranes deformation enters the in-plane membrane stretching mode, the Schwering-type solution becomes applicable. To extract the in-tensile stiffness from **Eq. 1**, the indentation depth has to be large so the second term therein dominates. The two-dimensional stiffness $E_{2D}$ and Young's modulus $Y$ extracted from our MD the simulation results are summarized in **Fig. 3a** as a function of the monovacancy concentration $c$. It can be clearly seen that the stiffness does show an increase with $c$ at low defect concentration, with its peak appearing at $c \sim 0.1\%$. This characteristic concentration is close to the value in the experimental report.[5] This counter-intuitive defect-induced stiffening can be explained by the combined effects from both geometrical stiffening under nanoindentation that is significant at low vacancy concentration and in-plane softening that dominates at the high-concentration limit where the geometric effect is saturated.[4]

To illustrate this combined effect, an analytical model is now developed. We consider the distorted membrane as a conical membrane experiencing an indentation force $f$ at the center, as illustrated in **Fig. 4**. With a swelling-induced displacement $d_0$ in the load-free condition that increases with $c$ (**Fig. S4**), analysis on the geometry leads to a relation between the in-plane stretch $\Delta l/l$ of graphene membrane and the indentation depth $\Delta d$. On the other hand, the indentation force $f$ can be related to the in-plane tension $\sigma_{2D}$ as $\sigma_{2D} = f/2\pi a t \sin\theta$ from the equation of equilibrium, where $t$ is the thickness of membrane and $\theta$ is the tilt angle between stretched graphene and the reference plane of supporting substrate. Combining these results, one finally can predict the defect-induced increase in the Young's modulus as

$$\Delta E_{2D} = \frac{\pi t Y a^3 \left(a^2 - d_0^2\right)}{\left(a^2 + d_0^2\right)^{5/2} \left[\frac{1}{2}\ln\left(a^2 + d_0^2\right) - \ln\left(R\arctan\frac{d_0}{a}\right) - 1\right]} \tag{2}$$

As shown in **Fig. 4b**, predictions from **Eq. 2** clearly demonstrate the geometrical effect in our explanation for the MD simulation results, although quantitative agreement is difficult due to the simplification in the model geometry.

From **Fig. 3a**, we notice that the amplitude of stiffening (~10 %) from our MD simulations is not as significant as the measured value (~100 %) in experiments. The difference can be explained by the limited spatial dimension in the setup of MD simulations here. As discussed in our previous study,[14] the deformation regime of a supported membrane changes from bending- to tension-dominated as the span of membrane increases, while in experiments the size of supported graphene is usually a few microns, and the membrane assumption of **Eq. 1** holds only at latter regime. To overcome the size-limitation in MD simulations, the simplification we make in the model analysis, and further demonstrate the validity of our discussion above, we conduct continuum level analysis that is intrinsically scale-free and depends on the geometry of model setup only. We perform finite element methods (FEM) based simulations, where both the in-plane softening and areal expansion effects are explicitly included. To simulate the supported graphene membrane, we use an axisymmetric shell model with Young's modulus calculated from the MD simulations and a Poisson's ratio $\nu = 0.3$,[14, 22] where both in-plane stretching and out-of-plane bending are considered with their relative contributions determined by the ratio between the lateral span of membrane and its thin-shell thickness.

In practice, we parameterize the model by numerically fitting a linear elastic constitutive relation to our MD simulations results of tensile tests (**Fig. 1a**). For the indentation tests, a concentrated force is loaded in the center of the clamped membrane. From the relation between indentation load and depth, we calculate the indentation stiffness $E_{2D}$ and in-plane Young's modulus $Y$ from **Eq. 1**. For the latter, a thickness of $t = 0.34$ nm is assumed for graphene. A circular graphene sheet is deposited onto a micropore with a pore diameter of 2 μm, which is consistent with the value in the experimental setup of nanoindentation tests but much larger than that in the MD simulations.[5] A prestrain is then applied to the membrane to simulate the vacancy-induced areal expansion that leads to significant out-of-plane buckling to release the prestress, according to our MD simulation results (**Fig. 1b**). The reduced in-plane Young's modulus of graphene with mono-vacancies is described by the values from our MD simulated in-plane tensile tests (**Fig. 1**). This result, as shown in **Fig. 3b**, clearly supports the argument of expansion-induced stiffening due to a prominent geometrical effect. Quantitatively, the stiffening ratio is above 100%, which is also close to the experimental measurements of graphene membrane with a similar size.

Based on these results, we conclude that the observed graphene stiffening at finite concentration of vacancies can be explained by two competing effects from the irradiation-introduced vacancies – the reduction in the in-plane stiffness and areal expansion that leads to geometrical distortion out of the plane, which stiffens the elastic response under nanoindentation as a result. The thermal wrinkles may play a role in renormalize the mechanical response of soft membranes such as graphene, however, this effect is not necessary to explain the observed vacancy-induced stiffening effect under nanoindentation. Actually, the role of thermal fluctuation, although being discussed in a number of theoretical studies,[23] still awaits validation and more details discussion based on direct evidences from experimental data. For example, the reported tensile stiffness of single- and polycrystalline graphene from nanoindentation tests and **Eq. 1**, where thermal effects are excluded, are found to be very consistent with theoretical predictions and measurement for the in-plane stiffness of graphite.[7-10] To further clarify our findings, we carry out FEM calculations where the in-plane softening is included while the areal expansion is not. The results indicate that as the concentration of vacancies increases, both the in-plane stiffness and Young's modulus deduced from nanoindentation tests decrease, consistently validating our proposed mechanism. In addition to elucidating the mechanism of defect-induced stiffening effect under nanoindentation, the findings here also suggest that the geometrical effect introduced due to defect-engineering could become a practical way to render the applications of 2D materials.[2, 24]

Moreover, the study here also raise the attention on the applicability of **Eq. 1** in measuring mechanical properties of 2D materials, where significant out-of-plane displacement and deformation are expected before and during the indentation. The pretension or prestrain characterized in AFM measurements is extracted from the first linear term in **Eq. 1**, from which the experimental data is fitted,[5-7] which could be accumulated in the membrane during the preparation process, or the adhesion between graphene and substrate.[5, 25] However, the membrane under nanoindentation is not flat. In recent experimental work where graphene is supported by porous substrate with pore size of ~1 μm, the out-of-plane displacement of load-free graphene membrane is reported to be 2.5 and 5-7 nm for pristine and defective graphene, respectively.[5-7] According to our analysis, the geometrical effect is expected to be significant and thus the applicability of **Eq. 1**, especially the linear term that dominates at small-displacement range, should be carefully assessed to extract meaningful material parameters from the measurements. A direct measurement of the morphology of suspended graphene before and after the indentation load is applied could also provide some insights in resolving this issue.

## Acknowledgment

This work was supported by the National Natural Science Foundation of China through Grant 11222217, and the State Key Laboratory of Mechanics and Control of Mechanical Structures, Nanjing University of Aeronautics and Astronautics, through Grant No. MCMS-0414G01. The computation was performed on the Explorer 100 cluster system at Tsinghua National Laboratory for Information Science and Technology.

**Figures and Captions**

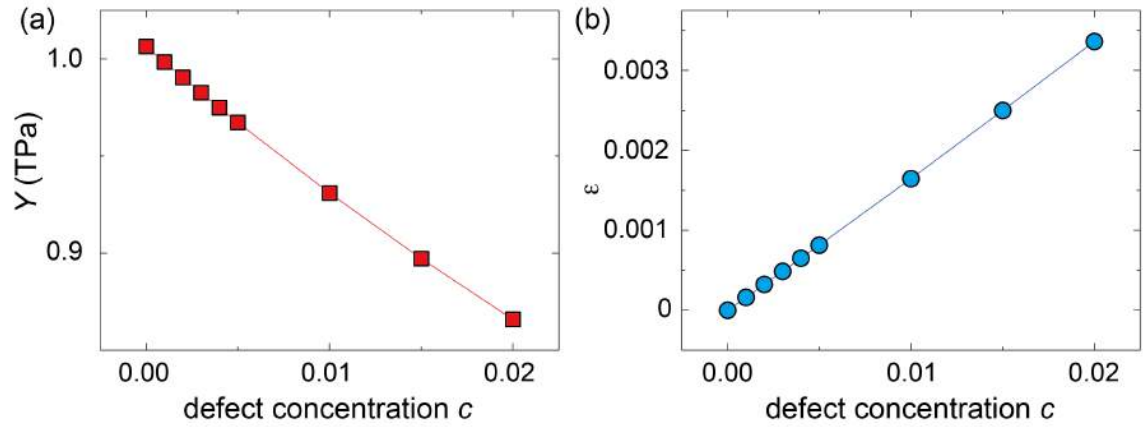

**Figure 1.** The reduction in the in-plane Young's modulus $Y$ as a function of the vacancy concentration $c$. (b) In-plane line expansion strain $\varepsilon$ caused by the presence of vacancies.

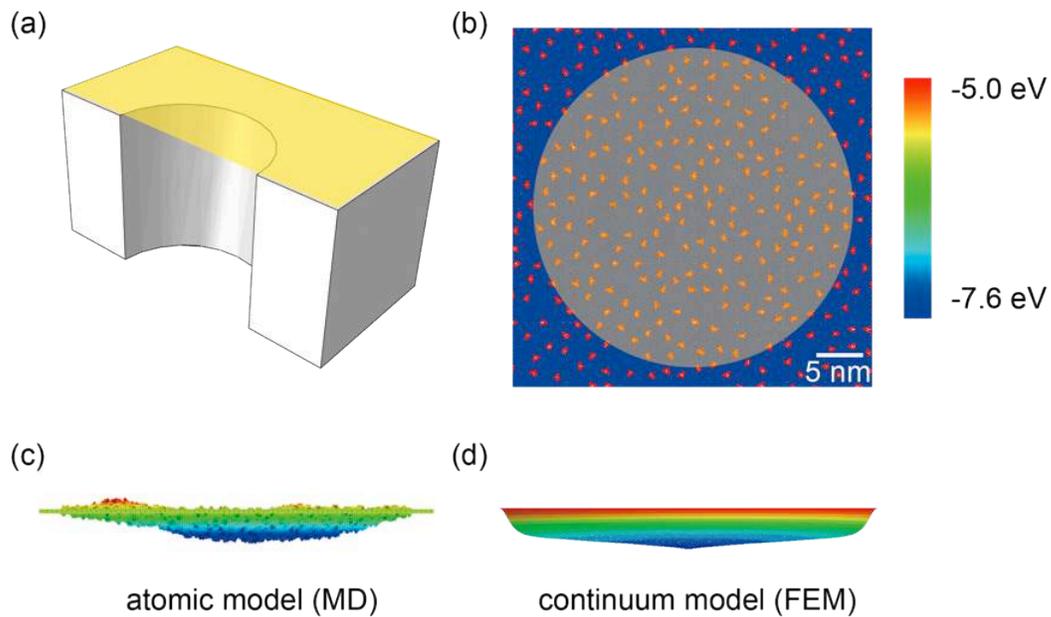

**Figure 2.** The illustration of the MD and FEM simulation setups. (a) The indentation setup with defective graphene supported by a porous substrate. (b) The distribution of mono-vacancies in the supported graphene membrane. The colors depict the potential energies of atoms. (c, d) The out-of-plane distortion of graphene membrane caused by the vacancy-induced expansion. The amplitude out-of-plane displacement is magnified by 5 times for illustration, and also depicted by colors.

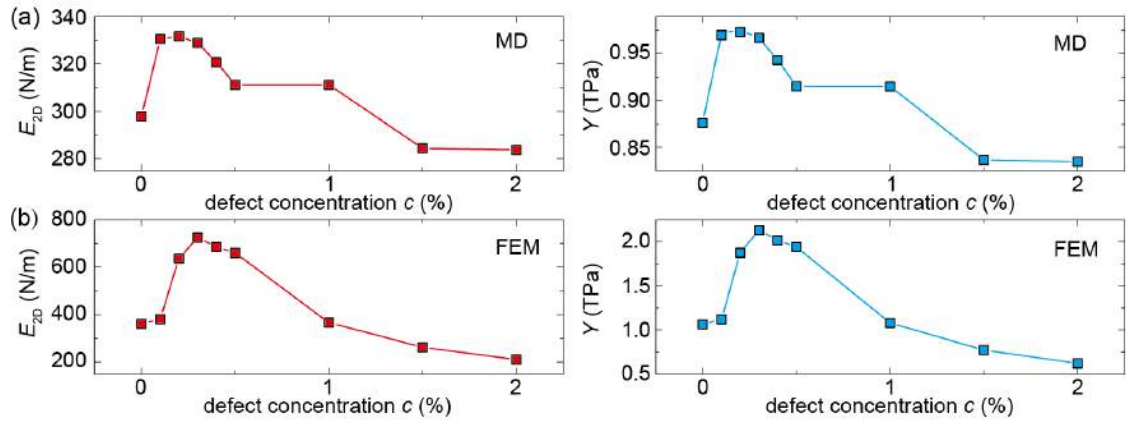

**Figure 3.** The in-plane stiffness $E_{2D}$ and Young's modulus $Y$ obtained by MD and FEM simulated nanoindentation tests, as a function of the concentration of vacancies $c$.

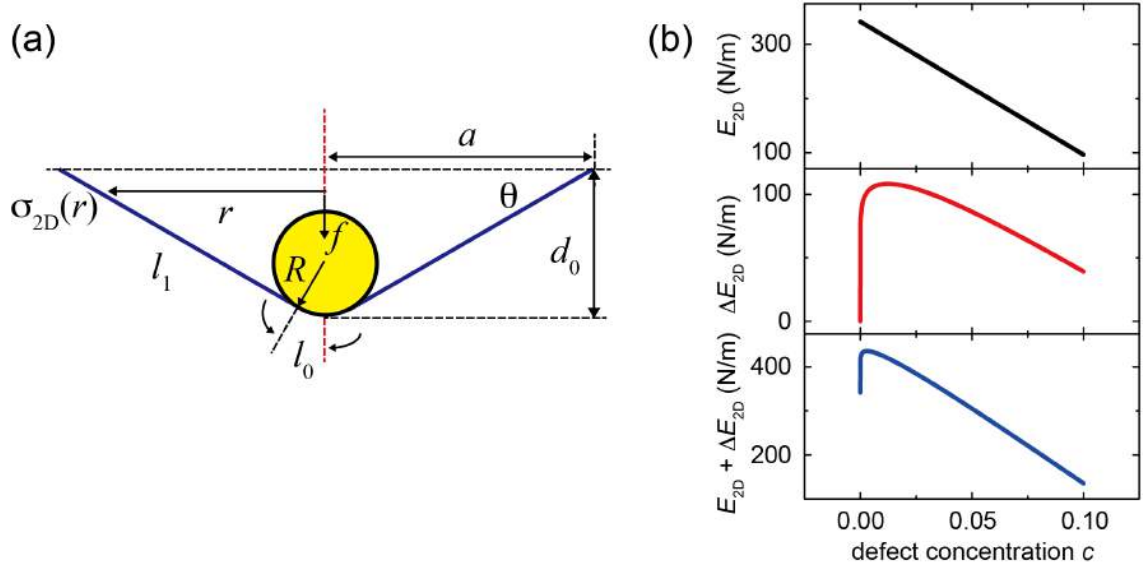

**Figure 4.** (a) An analytic membrane model of nanoindentation showing the geometric effect in defining the mechanical response of suspended graphene with a conic shape. (b) The in-plane stiffness $E_{2D}$ of defective graphene, the correction $\Delta E_{2D}$ introduced by the geometric effect that is calculated from **Eq. 2**, and the summation of these two contributions.